\documentclass[12pt,preprint]{aastex}

\slugcomment{Version: \today}

\shorttitle{EX~Lupi in Extreme Outburst}
\shortauthors{Aspin et al.}

\begin{document}

\title{THE 2008 EXTREME OUTBURST OF THE YOUNG ERUPTIVE VARIABLE STAR EX~LUPI}

\author{
Colin~Aspin\altaffilmark{1},
Bo~Reipurth\altaffilmark{1},
Gregory~J.~Herczeg\altaffilmark{2},
Peter~Capak\altaffilmark{3}}

\altaffiltext{1}{Institute for Astronomy, University of Hawai'i at
  Manoa, 640 N. A'ohoku Place, Hilo, HI 96720 \\ {\it
    (caa@ifa.hawaii.edu, reipurth@ifa.hawaii.edu)}}

\altaffiltext{2}{Max-Planck-Institut f\"{u}r Extraterrestriche Physik,
  Giessenbachstrasse, 85748 Garching, Germany\\ {\it
    (gregoryh@mpe.mpg.de)}}

\altaffiltext{3}{California Institute of Technology, MC 105-24, 1200
  East California Boulevard, Pasadena, CA 91125 \\ {\it
    (capak@me.com)}}

\begin{abstract} 
  In early 2008, the young low-mass star EX~Lupi, the prototype of the
  EXor class of eruptive variables, optically brightened by over five
  magnitudes for a period of 7 months.  The previous time a change of
  such amplitude had been observed in EX~Lup was over 50 years ago.
  In this Letter we present new optical and near-IR high resolution
  spectroscopy of EX~Lup during the 2008 outburst.  We investigate the
  physical characteristics of the outburst both soon after it began
  and some four months later, and consider the energetics and
  kinematics observed.  Emission line strengths, widths, and profiles
  changed significantly between the two observations.  Also, modeling
  of the 2.2935~$\mu$m CO overtone bandhead emission suggests that an
  inner gap in the circumstellar gas disk around the star may be
  present and it is from the inner edge of the gas disk that the CO
  overtone emission probably arises.  We derive a mass accretion
  luminosity and rate during the extreme outburst of
  $\sim$2$\pm$0.5~L$_{\odot}$ and
  $\sim$2$\pm$0.5$\times$10$^{-7}$~M$_{\odot}$~yr$^{-1}$,
  respectively, which suggests that this outburst was indeed one of
  the strongest witnessed in EX~Lup, yet not as strong as those
  observed in FU~Orionis stars.  \end{abstract}

\keywords{stars: individual (EX~Lupi) --- circumstellar matter ---
  stars: formation --- stars: variables: T Tauri, Herbig Ae/Be --- 
  accretion, accretion disks}

\section{INTRODUCTION}

Eruptive outbursts in young low-mass stars are scarce events that are
thought to be the result of increased accretion through a
circumstellar disk (Hartmann \& Kenyon 1996).  EX~Lupi is the
prototype of a class of young stars (termed EXors) showing such
outbursts (Herbig 1977, 1989, 2007).  It has been observed to exhibit
sporadic brightness increases of up to a few magnitudes which
typically last a few weeks before fading back to a quiescent state.
Such a 'characteristic outburst' occurred in 1993 and was studied
spectroscopically by Lehmann, Reipurth, \& Brandner (1995) and Herbig
et al. (2001).  In contrast to these smaller characteristic outbursts,
EX~Lup was observed in 1955 to exhibit a larger outburst resulting in
an optical brightness increase of $\sim$5 magnitudes. This event,
which we herein term an 'extreme outburst', was until recently unique
in the photometric annals of EX~Lup.  However, in early 2008 a second
such event took place.  Since it is unclear whether extreme outbursts
are initiated by the same physical mechanism as their smaller more
frequent counterparts, the advent of the 2008 eruption presented us
with the opportunity to perform a detailed observational study of an
extreme outburst.  Below, we present the results of this investigation
utilizing high spectral resolution optical (Keck~I/HIRES) and near-IR
(Keck~II/NIRSPEC) observations.

\section{OBSERVATIONS AND DATA REDUCTION}

Observations of EX~Lup (16:03:05.5,~--40:18:25,~J2000) were made on
the Keck~I and II 10-meter telescopes on UT 2008 January 23 (J23),
2008 February 23 (F23), and 2008 May 23 (M23).  Two instruments were
used: HIRES (Vogt et al. 1994) on Keck~I and NIRSPEC on Keck~II
(McLean et al. 1995).  Optical high-resolution spectra were obtained
with HIRES on J23 and M23.  On J23, the HIRESr configuration was used
with the KV380 filter and a slit width of 0$\farcs$861 resulting in a
spectral resolution of R$\sim$48,000 (i.e. a velocity resolution of
$\sim$7~km~s$^{-1}$) and a wavelength coverage of
$\sim$3850--8350~\AA.  On M23, the HIRESb configuration was used with
the WG335 filter and the same slit resulting in a similar spectral
resolution and a wavelength coverage of $\sim$3500--6300~\AA.  We are
greatly indebted to G.H.~Herbig for making the J23 HIRES spectra
available.  Near-IR (NIR) high-resolution spectra (R$\sim$18,000,
velocity resolution $\sim$17~km~s$^{-1}$) were obtained with NIRSPEC
on F23 and M23.

\section{RESULTS}

\subsection{Optical Light Curve}

\begin{figure*}[t] 
\begin{center}
\includegraphics*[angle=0,scale=1.2]{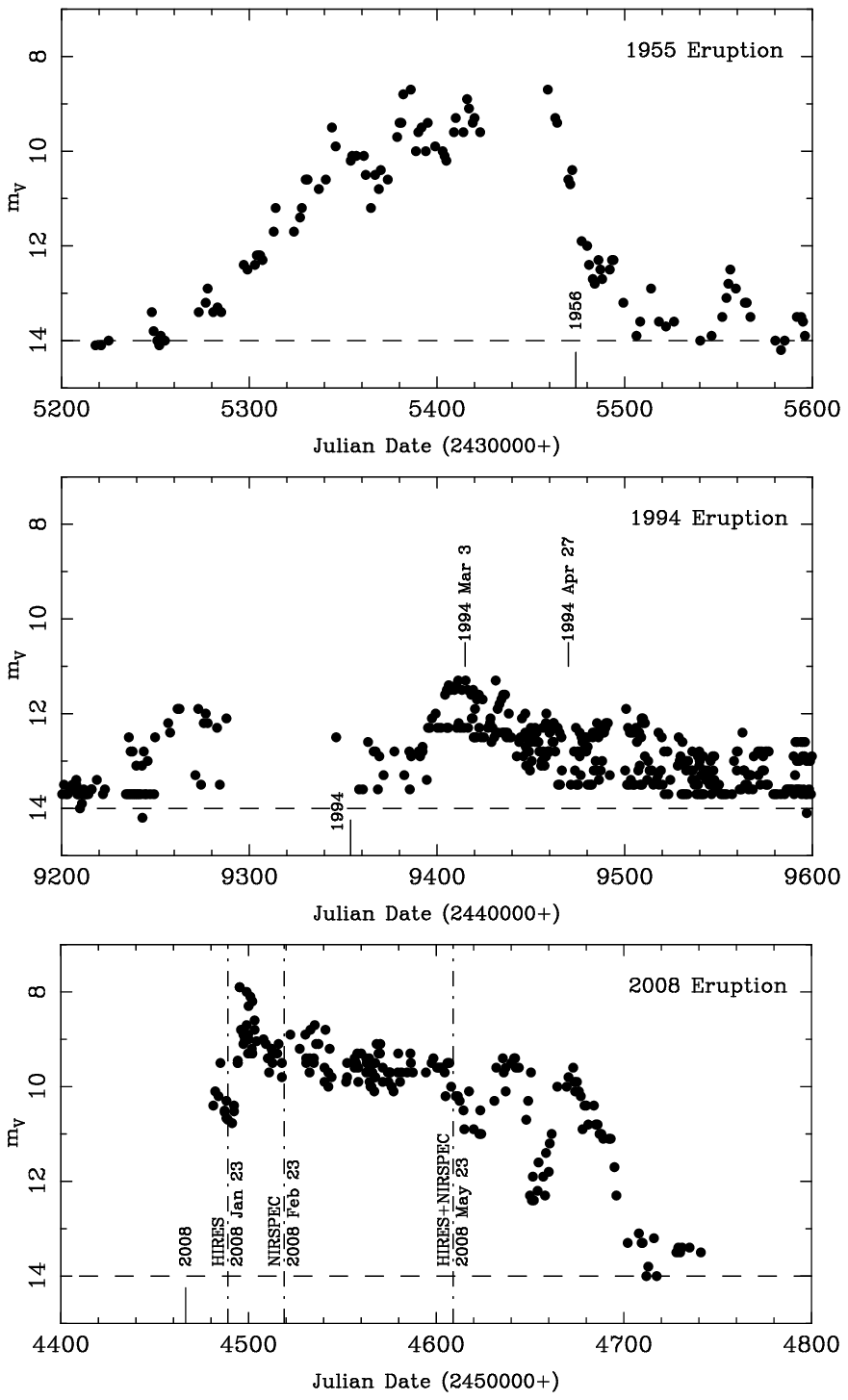} 

\caption{Optical light curves during the 1955 (top), 1993 (middle),
  and 2008 (bottom) outbursts.  All plots cover 400 days starting from
  shortly before the outbursts began.  The mean 'quiescent' optical
  brightness of EX~Lup is indicated by the horizontal dashed lines.
  All data points are taken from the AAVSO database.  The vertical
  dot-dashed lines in the bottom plot indicate the times when the
  HIRES and NIRSPEC spectroscopic observations were acquired.  The
  dates shown in the middle plot are those when the Lehmann et al.
  (1995) optical spectra were taken. \label{lc}}

\end{center}
\end{figure*}

Fig.~\ref{lc} shows the optical light curves of EX~Lup over the 1955
(top), 1993 (middle), and 2008 (bottom) outbursts, taken from the
AAVSO database.  The 2008 outburst plot shows that the J23 HIRES
observations were acquired soon after the extreme outburst occurred,
some 8~days after we had been notified of the event (A.~Jones, private
communication to G.H.~Herbig) and while the source was still in a
rapid brightening phase with m$_V\sim$10.5 or $\sim$3.5 magnitudes
brighter than in quiescence (m$_{V}\sim$14).  Our first NIRSPEC
observations were taken one month later when EX~Lup had passed its
peak brightness of m$_{V}\sim$8 and had m$_{V}\sim$9.5.  Our second
set of HIRES and NIRSPEC observations were taken simultaneously three
months later (M23), during the relatively constant 'plateau phase' of
the outburst which lasted $\sim$170~days.  At this time EX~Lup had
m$_{V}\sim$10.2 and was close to the start of a period of instability,
about two months prior to the sharp decline back to quiescence.  We
note that both the 1955 and 2008 extreme outbursts lasted
approximately the same length of time although they are somewhat
different in behavior; the 1955 eruption started slowly and built up
to a maximum (after $\sim$160~days) while the 2008 event appeared to
have a more rapid rise to maximum.  The light curve suggests that the
1993 event was a smaller characteristic outburst.

\subsection{HIRES Spectra}

\begin{figure*}[ht] 
\begin{center}
  \includegraphics*[angle=0,scale=1.0]{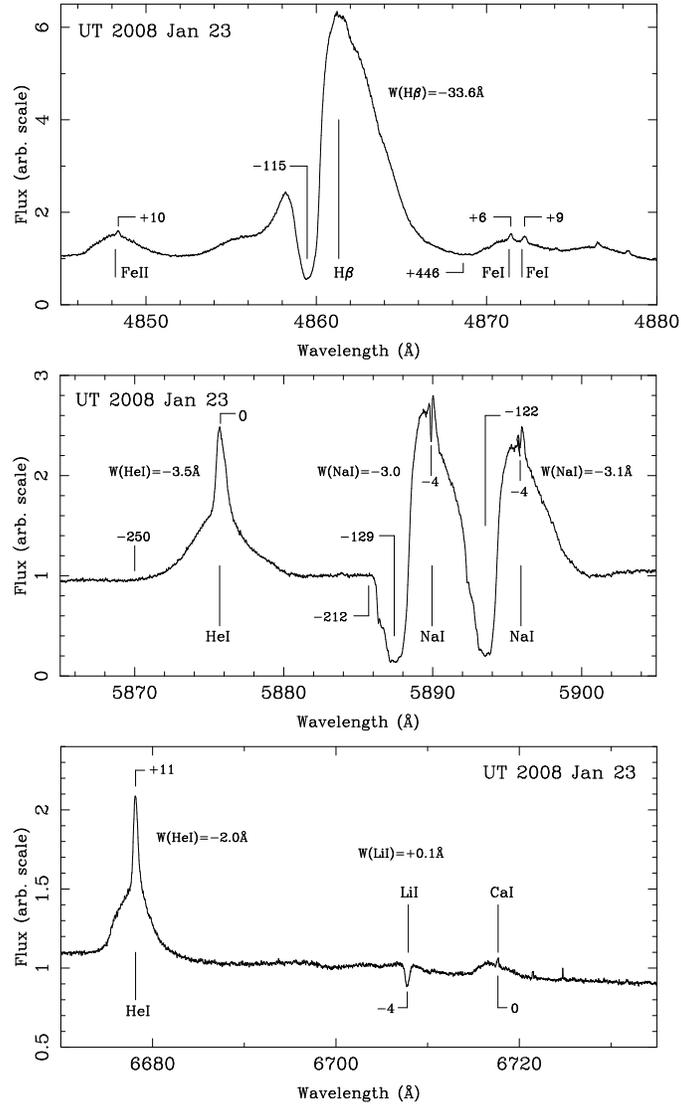} 

  \caption{Optical HIRES spectra taken on UT 2008 January 23 close to
    the start of the outburst.  Three wavelength regions are shown
    containing H$\beta$ (top), \ion{He}{1} and \ion{Na}{1} (middle),
    and \ion{He}{1} and \ion{Li}{1} (bottom).  Heliocentric radial
    velocities are given for the major features of the atomic lines.
    Equivalent widths (W) of the important features are shown.
    \label{hires1}}

\end{center}
\end{figure*}

In Fig.~\ref{hires1} we show three wavelength regions from the J23
HIRES spectrum, selected to cover important atomic features,
specifically, H$\beta$ (4861.3~\AA\ top), \ion{He}{1} (5875.7~\AA) and
\ion{Na}{1} (5890.0 and 5895.9~\AA\ middle), and \ion{He}{1}
(6678.2~\AA) and \ion{Li}{1} (6707.8~\AA\ bottom).  All but
\ion{Li}{1} are seen in emission (which is seen in absorption).
\ion{Na}{1} and H$\beta$ both possess blue-shifted absorption, with
the emission lines being broad (full width 10\% intensity
$\sim$750~km~s$^{-1}$) and the blue-shifted absorption removing flux
well below the continuum with a minimum at
$\sim$--120~km~s$^{-1}$.\footnotemark\footnotetext{All radial
  velocities are relative to the EX~Lup rest velocity.  EX~Lup has
  v$_{helio}$~=~+23, +28, and +2~km~s$^{-1}$, on J23, F23, and M23,
  respectively.}  The blue-shifted absorption on H$\beta$ appears
smooth with one broad minimum at --115~km~s$^{-1}$.  Similarly, the
blue-shifted absorption on the \ion{Na}{1} D lines have minima at
--129 and --122~km~s$^{-1}$, respectively.  \ion{Fe}{1}, FeII, and
\ion{He}{1} all exhibit composite profiles made up of both broad and
narrow emission components (BC and NC, respectively; Hamann \& Persson
1992).  The NC and BC peaks lie within $\sim$10~km~s$^{-1}$ of their
rest wavelength.

\begin{figure}[ht] 
\begin{center}
  \includegraphics*[angle=0,scale=1.0]{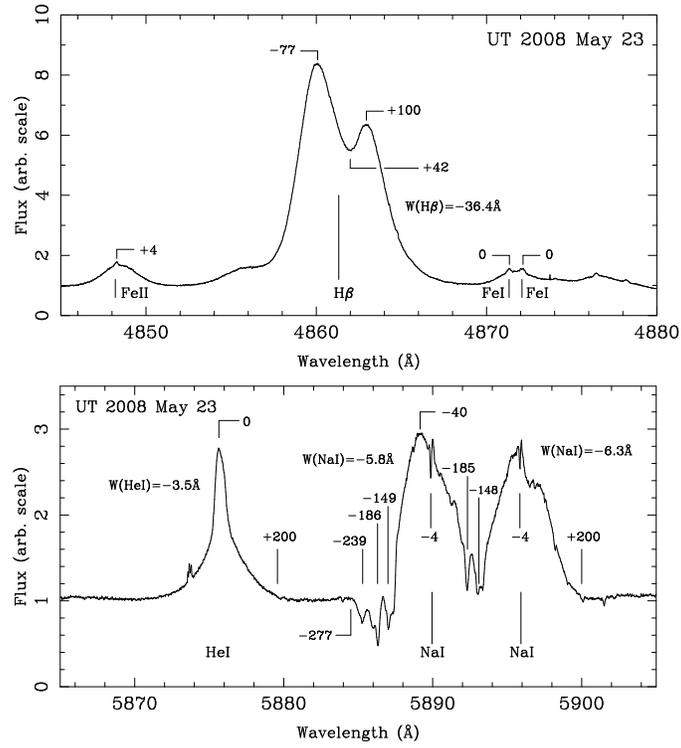} 

  \caption{Optical HIRES spectra taken on UT 2008 May 23 some 4 months
    after the outburst began.  Two of the three wavelength regions
    presented in Fig.~\ref{hires1} are shown here.  The HIRES echelle
    order containing the 6707~\AA\ \ion{Li}{1} line was not observed
    on this date due to the different echelle setting used.
    \label{hires2}}

\end{center}
\end{figure}

Four months after the first HIRES observation, on M23 (see
Fig.~\ref{hires2}), much of the same spectral structure was still
visible although some changes appeared to have
occurred\footnotemark\footnotetext{In the M23 observations,
  \ion{Li}{1} (6707.8~\AA) and \ion{He}{1} (6678.2~\AA) did not fall
  on the detector.}.  The \ion{He}{1} line still possessed NC and BC
features although the radial velocity offset of the NC from the line
rest wavelength had shifted to the blue.  Also, the H$\beta$ profile
had changed and exhibited a double peak at --77 and +100~km~s$^{-1}$
with an absorption minimum at +42~km~s$^{-1}$.  However, the width of
the H$\beta$ emission was similar to that found four months
previously.  The profiles of the \ion{Na}{1} D lines were also quite
different and became somewhat more complex.  Both lines remained in
emission and possessed blue-shifted absorption components although the
latter were much weaker than in the J23 spectra.  Additionally, the
blue-shifted absorptions showed radial velocity structure (see
Fig.~\ref{hires2}) extending to --277~km~s$^{-1}$ on M23 and
--212~km~s$^{-1}$ on J23.  Finally, superimposed on both of the broad
\ion{Na}{1} emission lines were narrow absorption features with a
radial velocity offset of --4~km~s$^{-1}$.

\begin{figure*}[ht] 
\begin{center}
  \includegraphics*[angle=0,scale=0.9]{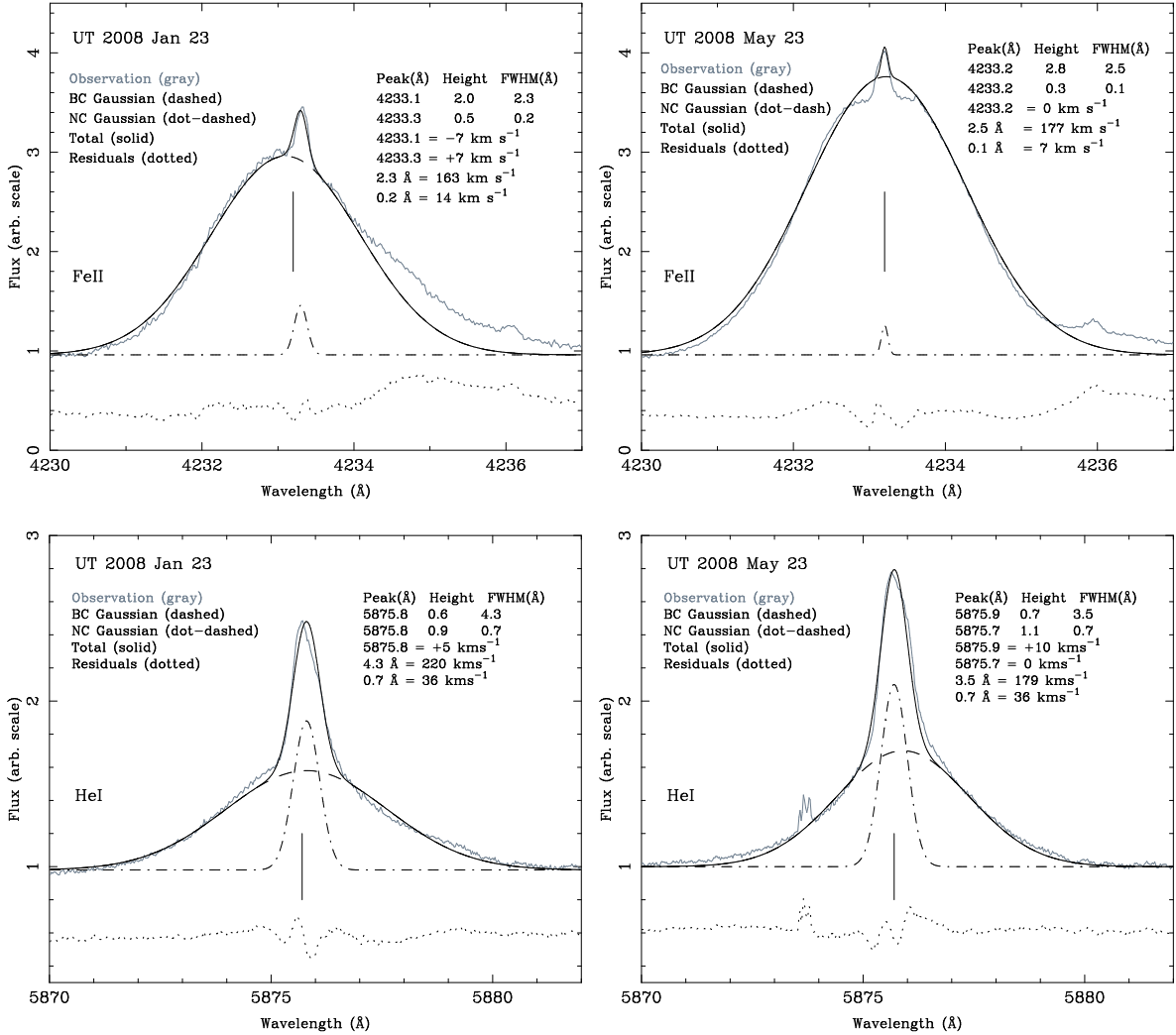} 

  \caption{HIRES spectra of the \ion{Fe}{2} line at 4233.2~\AA\ (top,
    grey lines) and the \ion{He}{1} line at 5876.1~\AA\ (bottom, grey
    lines) overlaid with model gaussian profiles.  The best fit
    parameters are given in the legends.  The NC gaussian is shown as
    dot-dashed lines while the BC gaussian is shown as dashed lines.
    The total model profile is the solid (black) lines and the
    residuals between the observed and model profiles are shown by the
    dotted lines.  The heliocentric rest wavelength of the line is
    shown as a vertical solid line. \label{heifeii}}

\end{center}
\end{figure*}

We have performed a best fit to the NC and BC structures present in
\ion{He}{1} and \ion{Fe}{2} using a two-component gaussian emission
profile model, iterating over three parameters for both gaussians,
specifically, the wavelength of peak emission ($\lambda_{peak}$), the
peak flux (F$_{peak}$), and the gaussian FWHM.  The results of this
procedure are shown in Fig.~\ref{heifeii}.  Between the two dates we
find that, {\it i)} for both lines $\lambda_{peak}$ of the NC and BC
gaussians remained within $\sim$10~km~s$^{-1}$ of the line rest
wavelength, {\it ii)} the \ion{He}{1} NC FHWM did not change
significantly, remaining at $\sim$36~km~s$^{-1}$, while the
\ion{Fe}{2} NC FWHM became smaller by a factor of 2 (14 and
7~km~s$^{-1}$) and at both times was significantly smaller than the
\ion{He}{1} NC FWHM, {\it iii)} the \ion{He}{1} BC FWHM declined (from
220 to 180~km~s$^{-1}$), while the \ion{Fe}{2} BC FWHM remained around
the same (at $\sim$170~km~s$^{-1}$), {\it iv)} the \ion{He}{1}
F$_{peak}$ of both the NC and BC increased slightly (by $\sim$20\%).
The \ion{He}{1} equivalent width (EW) on the two dates was
EW(\ion{He}{1})~=~--3.5~\AA\ (J23: EW$_{NC}$~=~--0.69~\AA,
EW$_{BC}$~=~--2.80~\AA, M23: EW$_{NC}$~=~--0.82~\AA,
EW$_{BC}$~=~--2.70~\AA).  Similarly, EW(\ion{Fe}{2}) on J23 was
--5.8~\AA\ (EW$_{NC}$~=~--0.21~\AA, EW$_{BC}$~=~--5.60~\AA) and on M23
was --7.9~\AA\ (EW$_{NC}$~=~--0.10~\AA, EW$_{BC}$~=~--7.80~\AA).

\subsection{NIRSPEC Spectra}

\begin{figure*}[ht] 
\begin{center}
  \includegraphics*[angle=0,scale=0.9]{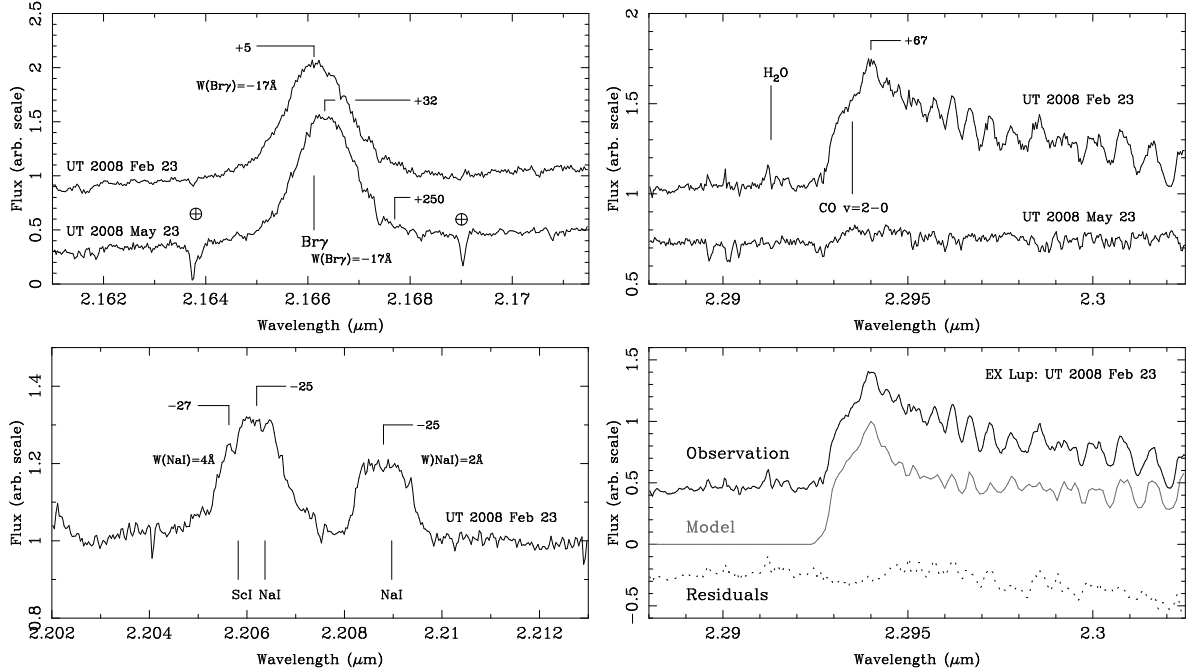} 

  \caption{NIRSPEC spectra taken during the extreme outburst of 2008.
    In two of the four plots, spectra taken at two different epochs
    are shown offset slightly for clarity (UT 2008 February 23 and May
    23).  Top-left is the Br$\gamma$ emission line at 2.166~$\mu$m.
    Bottom-left plot is the \ion{Na}{1} emission lines at 2.206 and
    2.209~$\mu$m.  Top-right plot is the v=2-0 CO overtone bandhead at
    2.294~$\mu$m. Bottom-right is the CO bandhead profile from UT 2008
    February 23 (black solid line) together with the best-fit disk
    emission model (grey solid line, offset for clarity).  The
    residuals between the observations and model are also shown
    (dotted line).  The best-fit model parameters are described in the
    text.  \label{nirspec1}}

\end{center}
\end{figure*}

Data for three spectral regions taken from the NIRSPEC observations
are shown in Fig.~\ref{nirspec1}.  Here, results from both observing
dates (F23 and M23) are plotted for the regions covering Br$\gamma$ at
2.166~$\mu$m (top-left) and the v=2--0 CO overtone bandhead at
2.2935~$\mu$m (top-right).  The bottom-left plot covers the
\ion{Na}{1} doublet lines at 2.206 and 2.208~$\mu$m from F23
only\footnotemark\footnotetext{The \ion{Na}{1} spectral region was not
  included in the M23 dataset.}.  The Br$\gamma$ line appears very
similar in J23 and M23 with a EW(Br$\gamma$) of --17~\AA. It is broad
(FWHM$\sim$220~km~s$^{-1}$ on F23, 190~km~s$^{-1}$ on M23), and is
slightly red-shifted with the $\lambda_{peak}$ emission offset by +5
and +32~km~s$^{-1}$ on F23 and M23, respectively.  The \ion{Na}{1}
doublet lines are also broad with FWHM of 177 and 155~km~s$^{-1}$ on
the one date they were observed.  The most striking change occurs in
the spectral region containing the 2.2935~$\mu$m v=2-0 CO overtone
bandhead.  In the data from F23, the bandhead is strongly in emission
with a EW(CO)~=~--21~\AA\footnotemark\footnotetext{calculated from
  2.2925 to 2.2975~$\mu$m.} and exhibits the characteristic signature
of being significantly velocity broadened (i.e. a 'blue-hump' is
present, see for example WL~16 in Carr et al. 1993).  In the data from
M23, however, the emission is only barely seen (EW(CO)~=~--4~\AA)
although velocity broadening of the bandhead still seems to be
present.  In the bottom-right plot of Fig.~\ref{nirspec1} we have
modeled the CO bandhead emission profile using the analysis from Dent
\& Geballe (1991) and Carr \& Tokunaga (1992) previously employed in
Aspin, Reipurth, \& Lehmann (1994) and Aspin \& Greene (2007).  The
details of the model are presented in the aforementioned papers and,
for brevity, here we merely present our findings.  The CO bandhead
emission is best fitted using a velocity profile originating in a
circumstellar disk.  This produces a good fit to the blue-hump and the
overall shape of the bandhead.  We investigated a wide range of the
model free parameters, specifically stellar mass (M$_{*}$) over the
range 0.1--2.0~M$_{\odot}$, the emission region temperature (T$_{CO}$)
from 1500 to 6000~K, the optical depth of CO ($\tau_{CO}$) from 0.01
to 5.0, and the inner (r$_{in}$) and outer (r$_{out}$) gas disk radii
between 1.0$\times$10$^{9}$ and 1.0$\times$10$^{11}$~m (i.e.
$\sim$1.4--140~R$_{*}$ or 0.007--0.7~AU). The best fit to the spectrum
gave values of M$_{*}$~=~0.6~M$_{\odot}$, T$_{CO}$~=~2500~K,
$\tau_{CO}$~$<$0.1, r$_{in}$~=~1.2$\times$10$^{10}$~m
($\sim$17~R$_{*}$ or 0.08~AU), and r$_{out}$~=~2.0$\times$10$^{10}$~m
($\sim$29~R$_{*}$ or 0.13~AU). Typical associated uncertainties on
these values should be considered to be $\sim$20\%.  For this value of
r$_{in}$ and with R$_{*}$~=~1.6~R$_{\odot}$ (as estimated by
Gras-Vel{\'a}zquez \& Ray 2005 and Sipos et al. 2010), gas at 2500~K
would require a heating source on the stellar surface of $\sim$8000~K
similar to what was found for surface accretion shocks in, for
example, V1647~Ori (Brice\~no et al. 2004).  The Keplerian velocity at
0.08~AU for M$_{*}$~=~0.6~M$_{\odot}$ is $\sim$80~km~s$^{-1}$.  Due to
the complex relationship between the model free parameters, the
uniqueness of the above best-fit model is unclear.  However, we
believe that our investigation of the aforementioned free-parameter
space results in values for the above physical properties that are
reasonable for this star.

\section{Discussion}

\subsection{Photometric variability}

The 2008 extreme outburst of EX~Lup appears to be unlike any other
observed in this source since 1955.  Both extreme outbursts exhibited
very large optical brightness increases (5--6 magnitudes) and,
although information about the earlier event is scarce they appear, at
least superficially, to be quite similar.  Between 1955 and 2008,
EX~Lup has undergone numerous smaller characteristic outbursts with
$\Delta$m$_{V}$ typically 1--3~magnitudes.  The durations of the
extreme and characteristic events appear very different; the former
last 200+ days whereas the latter last perhaps only a few weeks (see
the AAVSO archive for more details).  Photometrically, therefore,
there seems to be an inherent difference between extreme and
characteristic outbursts.

\subsection{Spectroscopic Variability}

The HIRES spectra of EX~Lup taken between 1998 and 2004 (presented in
Herbig 2007) and covering both the quiescent state and the
characteristic outburst phases, showed narrow emission features
together with an underlying atomic and molecular absorption spectrum
typical of an $\sim$M0 dwarf star.  During the 2008 extreme outburst,
however, the HIRES spectra appeared markedly different with no obvious
sign of an underlying photosphere.  Comparing our data taken during
the 2008 extreme outburst (shown in Figs.~\ref{hires1} and
\ref{hires2}) with previous observations we find that whereas Herbig
et al. (2001) found strong emission lines of H~I, \ion{He}{1},
\ion{He}{1}I, and \ion{Ca}{2} with no evidence for blue-shifted
absorption nor \ion{Fe}{1} and \ion{Fe}{2} lines, we observe both.
Also, the characteristic outburst phase spectrum from Lehmann,
Reipurth \& Brandner (1995) possessed \ion{H}{1}, \ion{He}{1},
\ion{He}{2}, \ion{Ca}{2}, lines together with weak \ion{Fe}{2} lines,
exhibiting inverse P~Cygni (IPC) profiles (red-shifted absorption) on
\ion{H}{1}, \ion{Ca}{2} and perhaps even \ion{Fe}{2} which disappeared
around 7 weeks later.  Herbig (1950) noted that IPC profiles were also
present in his 1949/1950 spectrograms taken several years before the
earlier extreme outburst.  Lehmann et al. (1995) interpreted these IPC
profiles as originating in cool infalling gas.  Our HIRES spectra,
which shows blue-shifted absorption, suggest that we are observing the
opposite, i.e. cool outflowing gas.  We speculate that perhaps it is
only during extreme outbursts that strong and dense enough
stellar/disk winds are generated to create the blue-shifted absorption
component.

Two-component (BC and NC) emission line profiles, similar to the ones
seen in our 2008 EX~Lup spectra, were previously reported on the
stronger \ion{Fe}{2} and Ti~II lines in EX~Lup in data taken in June
1998 (Herbig 2007).  Another young T~Tauri star to show such structure
is DR~Tau (Beristain, Edwards, \& Kwan 1998).  The interpretation
suggested for this source was that the NC of the lines arose in a
compact region close to the stellar surface (and hence not
significantly rotationally broadened) as in a post-shock region at the
base of an accretion flow.  The BC was considered to originate in a
region with considerable radial velocity structure such as in a
magnetospheric accretion funnel or in the hot inner region of the
circumstellar disk.  The different excitation requirements for
\ion{He}{1} (E$_{upper}$=23~eV) and \ion{Fe}{2} (E$_{upper}$=5--6~eV)
suggests that these lines are created in different physical locations,
making the similarity of structure somewhat peculiar.  Any model
proposed to explain the formation of these lines and the creation of
the NC+BC structure would need to explain this energy discrepancy
together with the temporal changes observed between J23 and M23 i.e.
{\it i)} a decline in the strength (i.e. EW) of the \ion{Fe}{2} NC by
a factor 2 coupled with a $\sim$40\% increase in the BC strength, {\it
  ii)} a $\sim$20\% increase in the strength of the \ion{He}{1} NC
coupled with an unchanged BC strength, and {\it iii)} the changes in
kinematic properties (e.g. FWHM, radial velocity offset) shown in
Fig.~\ref{heifeii}.

The changes in the blue-shifted absorption profiles between J23 and
M23 suggest that the wind from EX~Lup evolved between the two
observations.  The blue-shifted absorption features present on J23 and
seen in the H, Na, and Ca lines, varied somewhat differently for each
element.  Between J23 and M23, {\it i)} the blue-shifted H$\beta$
absorption moved redward and, in fact, became somewhat red-shifted,
{\it ii)} the blue-shifted \ion{Ca}{1} absorption disappeared, and
{\it iii)} the deep and broad single component blue-shifted absorption
on the \ion{Na}{1} lines became shallower and more complex in
structure, and extended to much higher radial velocities.  A
sophisticated radiative transfer code is required to model such
variations.

The changes in the CO overtone bandhead observed between J23 and M23,
together with the modeling presented above, suggest that the hot gas
in the inner regions of the accretion disk either cooled significantly
between the two dates, or was mostly accreted onto the star, or the UV
photoexcitation conditions changed (see Brittain, Najita, \& Carr
2009).  The unchanged Br$\gamma$ flux over this period implies strong
on-going accretion.

\subsection{Accretion Luminosity and Rate}

On J23, F23, and M23, EX~Lup had m$_V$ values of 10.5, 9.5, and 10.2,
respectively.  Juhasz et al. (2010, in preparation) measured the
K-band magnitude of EX~Lup to be m$_K$=7.5 on UT~2008~April~20.  On
that date EX~Lup had a V magnitude of $\sim$9.5 and hence a V--K=2.0.
If we assume this V--K value for F23 and M23 then on those dates m$_K$
was 7.5 and 8.2, respectively (a mean of 7.85).  The Br$\gamma$
emission line fluxes therefore decreased by 2$\times$ between J23 and
M23 since on both dates the EW of Br$\gamma$ was --17~\AA.  If the
optical continuum emission and the hydrogen line fluxes are indicative
of the mass accretion rate (e.g. Gullbring et al.  1998; Muzerolle,
Calvet, \& Hartmann 1998; Muzerolle, Hartmann, \& Calvet 1998) then
the accretion rate changed by only a factor 2 over this four month
period.

We can convert line and continuum luminosities to accretion luminosity
(L$_{acc}$) to estimate the mass accretion rate (\.M).  Following
Aspin, Beck, \& Reipurth (2008) and adopting m$_{K}$=7.85, a flux
corresponding to zeroth magnitude at 2.2~$\mu$m of
4.35$\times$10$^{-10}$~W~m$^{-2}$~$\mu$m$^{-1}$ (Tokunaga \& Vacca
2005), a visual extinction A$_V\sim$0, r$_{in}$=17~R$_{*}$ (see
above), L$_{*}$=0.5~L$_{\odot}$ (derived from T$_{*}$=3800~K and
R$_{*}$=1.6~R$_{\odot}$, again see above), and a distance of
140$\pm$20~pc (Hughes et al. 1993), we obtain values of
L$_{acc}$~1.6$\pm$0.6~L$_{\odot}$ and
\.M=1.5$\pm$0.5$\times$10$^{-7}$~M$_{\odot}$~ yr$^{-1}$.  Following
Valenti et al. (1993) and Gullbring et al. (1998), the change in
optical continuum emission in extreme outburst ($\Delta$V=4.5
magnitudes) is converted into L$_{acc}\sim$2.3~L$_{\odot}$ and
$\dot{M}\sim$2.1$\times$10$^{-7}$~M$_{\odot}$~yr$^{-1}$.  We therefore
adopt extreme outburst values of L$_{acc}$=2$\pm$0.5~L$_{\odot}$ and
\.M=2$\pm$0.5$\times$10$^{-7}$~M$_{\odot}$~ yr$^{-1}$.  Sipos et
al. (2009) estimated the quiescent accretion luminosity and rate of
EX~Lup to be L$_{acc}\sim$0.0037~L$_{\odot}$ and
$\dot{M}\sim$4.0$\times$10$^{-10}$~M$_{\odot}$~yr$^{-1}$ using
Pa$\beta$ line fluxes.  Since EX~Lup is veiled even in
quiescence\footnotemark\footnotetext{Herbig (2007) estimated an
  optical veiling of $\sim$1 when EX~Lup was faint.} we adopt
veiling-corrected values for quiescent L$_{acc}$ and \.M of
0.05$\pm$0.02 and 6$\pm$3$\times$10$^{-9}$, respectively.

The accretion luminosity and rate during the 2008 extreme outburst of
EX~Lup are therefore $\sim$40$\times$ larger than their quiescent
values.  However, these increases are still $\sim$50$\times$ smaller
than those found in FU~Orionis eruptions (Hartmann \& Kenyon 1996).

In conclusion, the 2008 extreme outburst of EX~Lup possessed a number
of features that distinguished it from the more common characteristic
outbursts and appears to have had more in common with the previous
extreme outburst that occurred in 1955.

\vspace{0.3cm}

{\bf Acknowledgments} 

We warmly thank G.H.~Herbig for providing access to the J23 HIRES data
on EX~Lupi.  We are indebted to A.~Jones for alerting us (via
G.H.~Herbig) to this new outburst of EX~Lup.  We also thanks A.~Juhasz
for providing the K-band magnitude used above.  Data were obtained at
the W.M. Keck Observatory, operated as a scientific partnership by the
California Institute of Technology, the University of California and
the National Aeronautics and Space Administration.  BR acknowledges
partial support from the NASA Astrobiology Institute (Cooperative
Agreement No. NNA04CC08A).

\clearpage

\begin{thebibliography}{}

\bibitem[Aspin \& Greene(2007)]{2007AJ....133..568A} Aspin, C. \&
  Greene, T.~P.\ 2007, \aj, 133, 568

\bibitem[Aspin et al.(1994)]{1994A&A...288..165A} Aspin, C., Reipurth,
  B., \& Lehmann, T.\ 1994, \aap, 288, 165

\bibitem[Aspin et al.(2008)]{2008AJ....135..423A} Aspin, C., Beck,
  T.~L., \& Reipurth, B.\ 2008, \aj, 135, 423

\bibitem[Beristain et al.(1998)]{1998ApJ...499..828B} Beristain, G.,
  Edwards, S., \& Kwan, J.\ 1998, \apj, 499, 828

\bibitem[Brice{\~n}o et al.(2004)]{2004ApJ...606L.123B} Brice{\~n}o,
  C., et al.\ 2004, \apjl, 606, L123

\bibitem[Brittain et al.(2009)]{2009ApJ...702...85B} Brittain, S.~D.,
  Najita, J.~R., \& Carr, J.~S.\ 2009, \apj, 702, 85

\bibitem[Carr \& Tokunaga(1992)]{1992ApJ...393L..67C} Carr, J.~S. \&
  Tokunaga, A.~T.\ 1992, \apjl, 393, L67

\bibitem[Carr et al.(1993)]{1993ApJ...411L..37C} Carr, J.~S.,
  Tokunaga, A.~T., Najita, J., Shu, F.~H., \& Glassgold, A.~E.\ 1993,
  \apjl, 411, L37

\bibitem[Dent \& Geballe(1991)]{1991A&A...252..775D} Dent, W.~R.~F. \&
  Geballe, T.~R.\ 1991, \aap, 252, 775

\bibitem[Gras-Vel{\'a}zquez \& Ray(2005)]{2005A&A...443..541G}
  Gras-Vel{\'a}zquez, {\'A}. \& Ray, T.~P.\ 2005, \aap, 443, 541

\bibitem[Gullbring et al.(1998)]{1998ApJ...492..323G} Gullbring, E.,
  Hartmann, L., Briceno, C., \& Calvet, N.\ 1998, \apj, 492, 323

\bibitem[Hamann \& Persson(1992)]{1992ApJS...82..247H} Hamann, F. \&
  Persson, S.~E.\ 1992, \apjs, 82, 247

\bibitem[Hartmann \& Kenyon(1996)]{1996ARA&A..34..207H} Hartmann, L.
  \& Kenyon, S.~J.\ 1996, \araa, 34

\bibitem[Herbig(1950)]{1950PASP...62..211H} Herbig, G.~H.\ 1950,
  \pasp, 62, 211

\bibitem[Herbig(1977)]{1977ApJ...217..693H} Herbig, G.~H.\ 1977, \apj,
  217, 693

\bibitem[Herbig(1989)]{1989lmsf.conf..233H} Herbig, G.~H.\ 1989,
  Workshop on {\it Low Mass Star Formation and Pre-main Sequence
    Objects}, ed. Bo~Reipurth, ESO-Garching, p.233

\bibitem[Herbig(2007)]{2007AJ....133.2679H} Herbig, G.~H.\ 2007, \aj,
  133, 2679

\bibitem[Herbig et al.(2001)]{2001PASP..113.1547H} Herbig, G.~H.,
  Aspin, C., Gilmore, A.~C., Imhoff, C.~L., \& Jones, A.~F.\ 2001,
  \pasp, 113, 1547

\bibitem[Hughes et al.(1993)]{1993AJ....105..571H} Hughes, J.,
  Hartigan, P., \& Clampitt, L.\ 1993, \aj, 105, 571

\bibitem[Johnson \& Apps(2009)]{2009ApJ...699..933J} Johnson, J.~A. \&
  Apps, K.\ 2009, \apj, 699, 933

\bibitem[Lehmann et al.(1995)]{1995A&A...300L...9L} Lehmann, T.,
  Reipurth, B., \& Brandner, W.\ 1995, \aap, 300, L9 [LRB95]

\bibitem[McLean et al.(1995)]{1995SPIE.2475..350M} McLean, I.~S.,
  Becklin, E.~E., Figer, D.~F., Larson, S., Liu, T., \& Graham, J.\
  1995, \procspie, 2475, 350

\bibitem[Muzerolle et al.(1998)]{1998ApJ...492..743M} Muzerolle, J.,
  Calvet, N., \& Hartmann, L.\ 1998, \apj, 492, 743

\bibitem[Muzerolle et al.(1998)]{1998AJ....116.2965M} Muzerolle, J.,
  Hartmann, L., \& Calvet, N.\ 1998, \aj, 116, 2965

\bibitem[Shortridge(1993)]{1993ASPC...52..219S} Shortridge, K.\ 1993,
  in {\it Astronomical Data Analysis Software and Systems II}, ed.
  R.~Hanisch, R.J.V.~Brissenden, \& J.~Barnes (San Francisco: ASP)
  p.219

\bibitem[Sipos et al.(2009)]{2009A&A...507..881S} Sipos, N.,
  {\'A}brah{\'a}m, P., Acosta-Pulido, J., Juh{\'a}sz, A.,
  K{\'o}sp{\'a}l, {\'A}., Kun, M., Mo{\'o}r, A., \& Setiawan, J.\
  2009, \aap, 507, 881

\bibitem[]{}Tokunaga, A. \& Vacca, W.~D.\ 2005, \pasp, 117, 1459

\bibitem[Valenti et al.(1993)]{1993AJ....106.2024V} Valenti, J.~A.,
  Basri, G., \& Johns, C.~M.\ 1993, \aj, 106, 2024

\bibitem[Vogt et al.(1994)]{1994SPIE.2198..362V} Vogt, S.~S., et al.\
  1994, \procspie, 2198, 362

\end{thebibliography}
\end{document}